\documentstyle[11pt]{article}

\newcommand{\be}{\begin{equation}}
\newcommand{\ee}{\end{equation}}
\newcommand{\bea}{\begin{eqnarray}}
\newcommand{\eea}{\end{eqnarray}}

\begin{document}

\begin{titlepage}

\vspace*{4cm}
\begin{center}

{\bf
SOLUTION OF THE SU(2) MANDELSTAM CONSTRAINTS
\\}
\vspace*{1cm}

{\bf N. J. Watson} \\
Centre de Physique Th\'eorique, C.N.R.S. - Luminy, Case 907, \\
F-13288 Marseille Cedex 9, France. \\
email: watson@cptsu1.univ-mrs.fr \\

\vspace*{1.5cm}
{\bf   Abstract  \\ }

\end{center}
It is shown how the Mandelstam constraints for an $SU(2)$ pure lattice
gauge theory with $3{\cal N}$ physical degrees of freedom may be solved
completely in terms of $3{\cal N}$ Wilson and Polyakov
loop variables and ${\cal N}-1$ gauge invariant discrete $\pm 1$
variables, thus enabling a manifestly gauge invariant formulation of the
theory.

\vspace*{1cm}

\noindent
November 1993

\noindent
CPT-93/P.2963

\vfill
\end{titlepage}

{\bf 1. Introduction}

\noindent
The idea that a gauge theory may be formulated entirely in terms of
gauge invariant variables is an appealing one, and has a long
though not particularly successful history
\cite{mandelstam1}\cite{polyakov1}\cite{makeenko}\cite{mig}\cite{polyakov2}.
For a pure gauge theory, i.e.
without external sources, the gauge invariant quantities are formed by taking
the trace of matrices $U_{\Gamma}$ given by the exponential of
the vector potential integrated around closed paths $\Gamma$:
\be\label{1.1}
{\rm Tr}\,U_{\Gamma}
=
{\rm Tr\,P\,exp}\oint_{\Gamma}A_{\mu}^{i}T^{i}dx^{\mu}
\ee
where for $SU(N)$
$A_{\mu}^{i}$, $i = 1,2\ldots N^{2}-1$, are the
vector potentials, the $T^{i}$ are the group generator matrices and P
denotes path ordering. Such variables are generally known as Wilson loops
in the lattice regularized versions of the theory.

One of the fundamental features of such approaches is the infinite
over-completeness of such loop variables. This implies the
existence of dependences among the loops, usually known as Mandelstam
constraints \cite{mandelstam2}.
These are complicated non-linear identities among
loops which intersect or touch at at least one point, and stem just
from the basic properties of $SU(N)$ matrices. The form of the
Mandelstam constraints for a given $SU(N)$ depends on $N$ (so that
going to the large-$N$ limit is not straightforward). For $N = 2$,
which is the simplest case, they may all be derived from the
fundamental identity\footnote{The corresponding identity for any given
$SU(N)$ may be easily derived from the Cayley-Hamilton theorem.}
for arbitrary $SU(2)$ matrices $U_{\alpha}$, $U_{\beta}$
\be\label{1.2}
{\rm Tr}\,U_{\alpha}U_{\beta}
-{\rm Tr}\,U_{\alpha}\,{\rm Tr}\,U_{\beta}
+ {\rm Tr}\,U_{\alpha}U_{\beta}^{-1}
=
0
\ee
A simple example of a Mandelstam constraint for $SU(2)$ Wilson loops
in the lattice versions of the
theory, in this case coming directly from eqn. (2),
is shown in fig. 1.

\noindent
\begin{picture}(425,140)(10,0)

\put(60,69){\line(0,1){21}}
\put(60,69){\line(1,0){39}}
\put(60,90){\line(1,0){41}}
\put(99,69){\line(0,-1){19}}
\put(101,71){\line(0,1){19}}
\put(101,71){\line(1,0){19}}
\put(99,50){\line(1,0){21}}
\put(120,50){\line(0,1){21}}

\put(145,70){\makebox(0,0){${\displaystyle - }$}}

\put(170,71){\framebox(39,19){}}
\put(211,50){\framebox(19,19){}}

\put(255,70){\makebox(0,0){${\displaystyle + }$}}

\put(280,70){\line(1,0){60}}
\put(280,70){\line(0,1){20}}
\put(280,90){\line(1,0){40}}
\put(320,50){\line(0,1){40}}
\put(320,50){\line(1,0){20}}
\put(340,50){\line(0,1){20}}

\put(363,70){\makebox(0,0)[l]{${\displaystyle = \hspace{15pt}0 }$}}

\put(10,10){\makebox(0,0)[l]{\footnotesize{
Fig. 1. A simple example of a Mandelstam constraint for Wilson
loops in $SU(2)$ lattice gauge theory.
}}}

\end{picture}

The number of independent physical degrees of freedom of an $SU(N)$ pure
gauge theory is determined by dividing out the gauge freedom from the
overall freedom of the system.
In a lattice formulation this number is regulated;
only such lattice theories will be considered here.
In the Lagrangian formulation \cite{wilson},
the total number of independent degrees of freedom of the space-time
lattice system is therefore given by
the dimension of the quotient space
$\otimes_{{\rm links}}SU(N)/\otimes_{{\rm sites}}SU(N)$.
In the Hamiltonian formulation \cite{kogsuss}, in which
a temporal gauge $A_{0}^{i} = 0$ is chosen and time remains continuous,
the number of independent degrees of freedom of the
spatial lattice system at a given time is similarly given by
dim$\{\otimes_{{\rm links}}SU(N)/\otimes_{{\rm sites}}SU(N)\}$.
In order to formulate either type of lattice
theory directly in terms of this number of
gauge invariant variables, it is necessary to be able to solve completely
the corresponding $SU(N)$ Mandelstam constraints in terms of a
complete independent set of this number of loop variables
(supplemented perhaps by some number of gauge invariant discrete
variables) so that {\em any} loop can then be expressed in terms of
this set.

For $SU(2)$, the two crucial identities required to solve
the Mandelstam constraints were given by Loll \cite{lollnpb1} (here
eqns. (\ref{2.13}) and (\ref{2.16})). However,
having derived the necessary identities, no attempt was then made to solve
them directly. Instead it was argued \cite{lollnpb1}\cite{lollconf}
that, assuming the non-existence of any
global constraints, it is sufficient to show the mutual independence
in any finite volume on the lattice of some specially chosen set of
Wilson loop variables, where the total number of loops in this set
equals the number of independent degrees of freedom of the system.
For example, for a two dimensional lattice
this set was taken to consist of
the one and two plaquette rectangular loops at each site.
Having demonstrated this independence,
then given that these variables altogether give the correct number of
degrees of freedom of the system and given the identities
necessary in principle
to solve completely the Mandelstam constraints, it was
argued that these variables should give a complete description of the
reduced configuration space of gauge invariant quantites of the system.
Further discussion was given in ref. \cite{lollnpb2}.

In this letter, it is shown how the Mandelstam constraints for
$SU(2)$ pure lattice gauge theory with
dim$\{\otimes_{{\rm links}}SU(2)/\otimes_{{\rm sites}}SU(2)\}=3{\cal N}$
variables may be solved
directly and completely in terms of $3{\cal N}$ continuous
Wilson and Polyakov
loop variables and ${\cal N}-1$ gauge invariant discrete
variables taking values $\pm 1$.
This enables a fully gauge invariant formulation of such
theories. The need for these discrete variables (which have nothing
to do with the $Z_{2}$ centre of the $SU(2)$ group) means that the
argument of Loll is incorrect, or, more precisely, incomplete.
Some possible practical consequences are briefly discussed.

{\bf 2. Solution of the SU(2) Mandelstam constraints}

\noindent
Consider a set of $m$ entirely arbitrary $SU(2)$ matrices, for
simplicity denoted generically now by $\alpha,\beta,\gamma\ldots$
rather than $U_{\alpha},U_{\beta},U_{\gamma}\ldots$. Then there exist
the following four identities, the derivations of which from eqn. (2) are
explained in ref. \cite{lollnpb1}:
\be\label{2.0}
{\rm Tr}(\alpha)
=
{\rm Tr}(\alpha^{-1})
\ee
Eqn. (\ref{2.0}) is due to the fact that the
representations of the $SU(2)$ generators are real.
\be\label{2.1}
{\rm Tr}(\alpha\beta)
=
{\rm Tr}(\alpha){\rm Tr}(\beta) - {\rm Tr}(\alpha\beta^{-1})
\ee
(eqn. (\ref{1.2}) trivially rearranged).
Using eqn. (\ref{2.1}), the trace of any product involving
matrices' inverses
can always be written in terms of traces with no inverses.
\bea\label{2.2}
{\rm Tr}(\alpha\beta\gamma)
&=&
-{\rm Tr}(\beta\alpha\gamma)
+{\rm Tr}(\alpha){\rm Tr}(\beta\gamma)
+{\rm Tr}(\beta){\rm Tr}(\gamma\alpha)
+{\rm Tr}(\gamma){\rm Tr}(\alpha\beta) \\ \nonumber
& &
-{\rm Tr}(\alpha){\rm Tr}(\beta){\rm Tr}(\gamma)
\eea
Using eqn. (\ref{2.2}), the trace of any product of matrices
in a given order can always be expressed in terms of the trace in any
other order and products of traces of products of fewer matrices.
\bea\label{2.3}
{\rm Tr}(\alpha\beta\gamma\delta)
&=&
\frac{1}{2}\Bigl(
{\rm Tr}(\alpha){\rm Tr}(\beta\gamma\delta)
+{\rm Tr}(\beta){\rm Tr}(\gamma\delta\alpha)
+{\rm Tr}(\gamma){\rm Tr}(\delta\alpha\beta)
+{\rm Tr}(\delta){\rm Tr}(\alpha\beta\gamma) \\ \nonumber
& &
-{\rm Tr}(\alpha\beta){\rm Tr}(\gamma){\rm Tr}(\delta)
-{\rm Tr}(\beta\gamma){\rm Tr}(\delta){\rm Tr}(\alpha) \\ \nonumber
& &
-{\rm Tr}(\gamma\delta){\rm Tr}(\alpha){\rm Tr}(\beta)
-{\rm Tr}(\delta\alpha){\rm Tr}(\beta){\rm Tr}(\gamma) \\ \nonumber
& &
+{\rm Tr}(\alpha\beta){\rm Tr}(\gamma\delta)
-{\rm Tr}(\alpha\gamma){\rm Tr}(\beta\delta)
+{\rm Tr}(\alpha\delta){\rm Tr}(\beta\gamma) \\ \nonumber
& &
+{\rm Tr}(\alpha){\rm Tr}(\beta){\rm Tr}(\gamma){\rm Tr}(\delta) \Bigr)
\eea
Using eqn. (\ref{2.3}), the trace of any product of four or more matrices
can always be expressed in terms of products of traces of one, two and
three matrices.

The identities eqns. (\ref{2.0})(\ref{2.1})(\ref{2.2})(\ref{2.3})
together enable
the trace of any product of matrices and their inverses from the given set
to be expressed in terms of the traces of products of one, two and
three matrices, which will be denoted generically by
${\rm Tr}(\alpha)$, ${\rm Tr}(\alpha\beta)$ and ${\rm Tr}(\alpha\beta\gamma)$.

However, the
${\rm Tr}(\alpha\beta)$ and ${\rm Tr}(\alpha\beta\gamma)$
are themselves constrained. To express these constraints, it is
convenient to introduce the so-called $L$ variables:\footnote{
The definitions of the $L$ variables used here differ slightly from
those used in ref. \cite{lollnpb1}.}
\bea\label{2.4}
L(\alpha)
&=&
{\textstyle\frac{1}{2}}{\rm Tr}(\alpha) \\
L(\alpha,\beta)
&=&
{\textstyle\frac{1}{4}}\Bigl(
-{\rm Tr}(\alpha\beta)
+{\rm Tr}(\alpha\beta^{-1})
\Bigr) \\
\label{2.5}
&=&
-{\textstyle\frac{1}{2}}{\rm Tr}(\alpha\beta)
+{\textstyle\frac{1}{4}}{\rm Tr}(\alpha){\rm Tr}(\beta) \\
L(\alpha,\beta,\gamma)
&=&
{\textstyle\frac{1}{16}}\Bigl(
-{\rm Tr}(\alpha\beta\gamma)
+{\rm Tr}(\alpha\beta^{-1}\gamma)
+{\rm Tr}(\alpha\beta\gamma^{-1})
-{\rm Tr}(\alpha\beta^{-1}\gamma^{-1}) \\ \nonumber
& &
+{\rm Tr}(\alpha\gamma\beta)
-{\rm Tr}(\alpha\gamma^{-1}\beta)
-{\rm Tr}(\alpha\gamma\beta^{-1})
+{\rm Tr}(\alpha\gamma^{-1}\beta^{-1}) \Bigr) \\
\label{2.6}
&=&
-{\textstyle\frac{1}{2}}{\rm Tr}(\alpha\beta\gamma)
+{\textstyle\frac{1}{4}}{\rm Tr}(\alpha){\rm Tr}(\beta\gamma)
+{\textstyle\frac{1}{4}}{\rm Tr}(\beta){\rm Tr}(\gamma\alpha)
+{\textstyle\frac{1}{4}}{\rm Tr}(\gamma){\rm Tr}(\alpha\beta) \\ \nonumber
& &
-{\textstyle\frac{1}{4}}{\rm Tr}(\alpha){\rm Tr}(\beta){\rm Tr}(\gamma)
\eea
These $L$ variables have simple geometrical interpretations:
writing the $SU(2)$ matrices in the form
\be\label{2.7}
\alpha
=
\alpha_{0}I + i\mbox{\boldmath $\alpha\cdot\sigma$}
\ee
where
$\mbox{\boldmath $\sigma$} = (\sigma_{1},\sigma_{2},\sigma_{3})$ are the
three Pauli matrices and the real numbers
$\alpha_{i}$, $i = 0,1,2,3$, satisfy
$\alpha_{0}^{2} + \mbox{\boldmath $\alpha\cdot\alpha$} = 1$
where
$\mbox{\boldmath $\alpha$} = (\alpha_{1},\alpha_{2},\alpha_{3})$
is a 3-vector, the $L$ variables may be written
\bea\label{2.8}
L(\alpha)
&=&
\alpha_{0} \\
\label{2.9}
L(\alpha,\beta)
&=&
\mbox{\boldmath $\alpha\cdot\beta$} \\
\label{2.10}
L(\alpha,\beta,\gamma)
&=&
\mbox{\boldmath $\alpha\times\beta\cdot\gamma$}
\eea
In this form, the properties of the $L$ variables under permutation
and inversion of their matrix arguments are particularly transparent
e.g. $L(\alpha,\beta) = L(\beta,\alpha)$.
Indeed, the identities eqns. (\ref{2.0})(\ref{2.1})(\ref{2.2})
expressed in terms of the $L$ variables become simply
$L(\alpha) = L(\alpha^{-1})$, $L(\alpha,\beta) = -L(\alpha,\beta^{-1})$
and $L(\alpha,\beta,\gamma) = -L(\beta,\alpha,\gamma)$
respectively. Also it can immediately be seen that
$L(\alpha,\beta)$ and $L(\alpha,\beta,\gamma)$ vanish
if any of their arguments is the unit matrix $I$.
It is clear that these $L$ variables are more convenient
quantities with which to work than the original traces.
The definitions eqns. (\ref{2.4})(\ref{2.5})(\ref{2.6})
may easily be inverted to give the
${\rm Tr}(\alpha)$, ${\rm Tr}(\alpha\beta)$ and
${\rm Tr}(\alpha\beta\gamma)$ in terms of the
$L(\alpha)$, $L(\alpha,\beta)$ and $L(\alpha,\beta,\gamma)$.

There then exist two further sets of identities, given originally in
ref. \cite{lollnpb1}. Firstly:
\bea\label{2.13}
L(\alpha,\beta,\gamma)^{2}
+L(\alpha,\beta)^{2}L(\gamma,\gamma)
+L(\beta,\gamma)^{2}L(\alpha,\alpha)
+L(\gamma,\alpha)^{2}L(\beta,\beta)
& & \\ \nonumber
-2L(\alpha,\beta)L(\beta,\gamma)L(\gamma,\alpha)
-L(\alpha,\alpha)L(\beta,\beta)L(\gamma,\gamma)
&=&
0
\eea
Using eqn. (\ref{2.13}), any $L(\alpha,\beta,\gamma)$, and hence any
${\rm Tr}(\alpha\beta\gamma)$, can always be expressed in terms of the
$L(\alpha,\beta)$, or the ${\rm Tr}(\alpha)$ and ${\rm Tr}(\alpha\beta)$,
up to a sign.
If the angle between two vectors \mbox{\boldmath $\alpha,\beta$} is written
$\theta_{\alpha\beta}$,
then an overall product of the lengths of the vectors in eqn.
(\ref{2.13}) may be factored out leaving the identity in the form
\be\label{2.14}
{\rm sin}^{2}\theta_{\alpha\beta}
{\rm cos}^{2}\theta_{(\alpha\times\beta)\gamma}
+{\rm cos}^{2}\theta_{\alpha\beta}
+{\rm cos}^{2}\theta_{\beta\gamma}
+{\rm cos}^{2}\theta_{\gamma\alpha}
-2\,{\rm cos}\,\theta_{\alpha\beta}
{\rm cos}\,\theta_{\beta\gamma}{\rm cos}\,\theta_{\gamma\alpha} -1
=
0
\ee
This identity expresses the fact that, given the angles
$\theta_{\alpha\beta}$, $\theta_{\beta\gamma}$, $\theta_{\gamma\alpha}$
between three vectors
\mbox{\boldmath $\alpha$},
\mbox{\boldmath $\beta$}, \mbox{\boldmath $\gamma$},
the relative orientation of the three vectors is specified up to a
possible
``reflection'' of, say, \mbox{\boldmath $\gamma$} in the plane formed
by \mbox{\boldmath $\alpha$} and \mbox{\boldmath $\beta$}. Thus there
are in general two solutions $\theta_{(\alpha\times\beta)\gamma}$ and
$\pi - \theta_{(\alpha\times\beta)\gamma}$ of the quadratic eqn.
(\ref{2.14}). This is illustrated in fig. 2(a). The only information
given by the $L(\alpha,\beta,\gamma)$ which is not supplied by the
$L(\alpha,\beta)$ is this relative reflectional orientation of
the sets of vectors
\mbox{\boldmath $\alpha$},
\mbox{\boldmath $\beta$}, \mbox{\boldmath $\gamma$}.
This information is given simply by the sign of each of the
$L(\alpha,\beta,\gamma)$, so that the solution of the equation of
constraint (\ref{2.13}) may be written
\be\label{2.15}
L(\alpha,\beta,\gamma)
=
s(\alpha,\beta,\gamma)\,|L(\alpha,\beta,\gamma)|
\ee
where the modulus $|L(\alpha,\beta,\gamma)|$
is given entirely by the $L(\alpha,\beta)$ as in eqn. (\ref{2.13}) and
the discrete variables $s(\alpha,\beta,\gamma)$ have value $+1$
if $\theta_{(\alpha\times\beta)\gamma} < \frac{1}{2}\pi$  and $-1$
if $\theta_{(\alpha\times\beta)\gamma} > \frac{1}{2}\pi$, so
specifying the relative reflectional orientations. This is illustrated
in figs. 2(b) and 2(c) which show schematically in
``plan view'' the angles
$\theta_{\alpha\beta}$, $\theta_{\beta\gamma}$, $\theta_{\gamma\alpha}$
and the two possible relative orientations of the vectors.
For $\theta_{(\alpha\times\beta)\gamma} = \frac{1}{2}\pi$ i.e.
\mbox{\boldmath $\alpha$},
\mbox{\boldmath $\beta$}, \mbox{\boldmath $\gamma$} coplanar and
$L(\alpha,\beta,\gamma) = 0$, or if any of the $\alpha,\beta,\gamma$
is the unit matrix so that again $L(\alpha,\beta,\gamma) = 0$,
$s(\alpha,\beta,\gamma)$ is undefined.
The $s(\alpha,\beta,\gamma)$ have the same properties under
permutation and inversion of their matrix arguments as the
$L(\alpha,\beta,\gamma)$.

\noindent
\begin{picture}(425,270)(10,-20)

\thicklines
\put(80,80){\vector(-1,2){33.33}}
\put(80,80){\vector(-1,4){20}}
\put(80,80){\vector(1,4){25}}
\put(80,80){\vector(4,3){40}}
\put(80,80){\vector(2,-1){30}}

\thinlines
\multiput(80,80)(-2,1){20}{\circle*{1}}
\multiput(120,110)(-2,1){18}{\circle*{1}}
\multiput(65.71,137.14)(-2,1){9}{\circle*{1}}

\put(109,196){\makebox(0,0){\boldmath $\alpha$}}
\put(56,176){\makebox(0,0){\boldmath $\beta$}}
\put(39.29,161.41){\makebox(0,0){\boldmath $\gamma$}}
\put(133.2,119.9){\makebox(0,0){\boldmath $\gamma$}}
\put(124.76,57.62){\makebox(0,0){\boldmath $\alpha\times\beta$}}

\put(200,150){\line(1,0){45}}
\put(200,150){\line(2,-3){30}}
\put(245,150){\line(-1,-3){15}}
\put(200,150){\circle*{3}}
\put(245,150){\circle*{3}}
\put(230,105){\circle*{3}}

\put(190,160){\makebox(0,0){\boldmath $\beta$}}
\put(255,160){\makebox(0,0){\boldmath $\alpha$}}
\put(230,90){\makebox(0,0){\boldmath $\gamma$}}
\put(205,125){\makebox(0,0){$\theta_{\beta\gamma}$}}
\put(250,125){\makebox(0,0){$\theta_{\gamma\alpha}$}}
\put(225,160){\makebox(0,0){$\theta_{\alpha\beta}$}}

\put(222.5,50){\makebox(0,0){\footnotesize{$s(\alpha,\beta,\gamma) = +1$}}}

\put(342.5,150){\line(1,0){45}}
\put(342.5,150){\line(2,3){30}}
\put(387.5,150){\line(-1,3){15}}
\put(342.5,150){\circle*{3}}
\put(387.5,150){\circle*{3}}
\put(372.5,195){\circle*{3}}

\put(332.5,140){\makebox(0,0){\boldmath $\beta$}}
\put(397.5,140){\makebox(0,0){\boldmath $\alpha$}}
\put(372.5,210){\makebox(0,0){\boldmath $\gamma$}}
\put(347.5,175){\makebox(0,0){$\theta_{\beta\gamma}$}}
\put(392.5,175){\makebox(0,0){$\theta_{\gamma\alpha}$}}
\put(367.5,140){\makebox(0,0){$\theta_{\alpha\beta}$}}

\put(365,50){\makebox(0,0){\footnotesize{$s(\alpha,\beta,\gamma) = -1$}}}

\put(80,30){\makebox(0,0){\footnotesize{(a)}}}
\put(222.5,30){\makebox(0,0){\footnotesize{(b)}}}
\put(365,30){\makebox(0,0){\footnotesize{(c)}}}

\put(10,10){\makebox(0,0)[l]{\footnotesize{
Fig. 2(a). The two possible orientations of $\gamma$
relative to $\alpha$, $\beta$ given the three angles
$\theta_{\alpha\beta},\theta_{\beta\gamma},\theta_{\gamma\alpha}$.
}}}
\put(10,0){\makebox(0,0)[l]{\footnotesize{
Fig. 2(b). The three angles
$\theta_{\alpha\beta},\theta_{\beta\gamma},\theta_{\gamma\alpha}$
represented schematically in ``plan view'' for $s(\alpha,\beta,\gamma) = +1$.
}}}
\put(10,-10){\makebox(0,0)[l]{\footnotesize{
Fig. 2(c). As 2(b) but for $s(\alpha,\beta,\gamma) = -1$.
}}}

\end{picture}

Secondly (and lastly):
\bea\label{2.16}
 L(\alpha,\beta)^{2}L(\gamma,\delta)^{2}
+L(\alpha,\gamma)^{2}L(\beta,\delta)^{2}
+L(\alpha,\delta)^{2}L(\beta,\gamma)^{2}
& & \\ \nonumber
-L(\alpha,\beta)^{2}L(\gamma,\gamma)L(\delta,\delta)
-L(\alpha,\gamma)^{2}L(\beta,\beta)L(\delta,\delta)
-L(\alpha,\delta)^{2}L(\beta,\beta)L(\gamma,\gamma)
& & \\ \nonumber
-L(\beta,\gamma)^{2}L(\alpha,\alpha)L(\delta,\delta)
-L(\beta,\delta)^{2}L(\alpha,\alpha)L(\gamma,\gamma)
-L(\gamma,\delta)^{2}L(\alpha,\alpha)L(\beta,\beta)
& & \\ \nonumber
+2L(\alpha,\alpha)L(\beta,\gamma)L(\gamma,\delta)L(\delta,\beta)
+2L(\beta,\beta)L(\alpha,\gamma)L(\gamma,\delta)L(\delta,\alpha)
& & \\ \nonumber
+2L(\gamma,\gamma)L(\alpha,\beta)L(\beta,\delta)L(\delta,\alpha)
+2L(\delta,\delta)L(\alpha,\beta)L(\beta,\gamma)L(\gamma,\alpha)
& & \\ \nonumber
-2L(\alpha,\beta)L(\beta,\gamma)L(\gamma,\delta)L(\delta,\alpha)
-2L(\alpha,\gamma)L(\beta,\gamma)L(\beta,\delta)L(\delta,\alpha)
& & \\ \nonumber
-2L(\alpha,\beta)L(\beta,\delta)L(\gamma,\delta)L(\gamma,\alpha)
+L(\alpha,\alpha)L(\beta,\beta)L(\gamma,\gamma)L(\delta,\delta)
&=&
0
\eea
Again, an overall
product of the lengths of the vectors in eqn. (\ref{2.16})
may be factored out leaving the identity entirely in terms
of the angles between the vectors. This identity expresses the
fact that, given four vectors
\mbox{\boldmath $\alpha$}, \mbox{\boldmath $\beta$},
\mbox{\boldmath $\gamma$}, \mbox{\boldmath $\delta$}, one of the six
angles between them, say $\theta_{\gamma\delta}$, may always be expressed
in terms of the other five,
$\theta_{\alpha\beta}$, $\theta_{\alpha\gamma}$,
$\theta_{\alpha\delta}$, $\theta_{\beta\gamma}$,
$\theta_{\beta\delta}$ in this case, up to a sign ambiguity:
although there are altogether four possible reflectional orientations
for the vectors
\mbox{\boldmath $\gamma$} and \mbox{\boldmath $\delta$}
in the plane formed by \mbox{\boldmath $\alpha$} and \mbox{\boldmath $\beta$},
only two of these give in general distinct values for $\theta_{\gamma\delta}$,
the other two being just overall reflections of these two orientations,
so that there are in general two solutions of this
quadratic equation of constraint.
This is illustrated in fig. 3 which shows
schematically the angles in the
four possible relative orientations of the vectors.

\noindent
\begin{picture}(425,260)(0,-20)

\put(30,150){\line(1,0){45}}
\put(30,150){\line(1,-6){5}}
\put(30,150){\line(2,-3){30}}
\put(75,150){\line(-1,-3){15}}
\put(75,150){\line(-4,-3){40}}
\multiput(35,120)(2.5,-1.5){10}{\circle*{1}}
\put(30,150){\circle*{3}}
\put(75,150){\circle*{3}}
\put(60,105){\circle*{3}}
\put(35,120){\circle*{3}}

\put(20,150){\makebox(0,0){\boldmath $\beta$}}
\put(85,150){\makebox(0,0){\boldmath $\alpha$}}
\put(65,95){\makebox(0,0){\boldmath $\gamma$}}
\put(25,110){\makebox(0,0){\boldmath $\delta$}}

\put(55,60){\makebox(0,0){\footnotesize{$s(\alpha,\beta,\gamma) = +1$}}}
\put(55,50){\makebox(0,0){\footnotesize{$s(\alpha,\beta,\delta) = +1$}}}

\put(135,150){\line(1,0){45}}
\put(135,150){\line(1,6){5}}
\put(135,150){\line(2,-3){30}}
\put(180,150){\line(-1,-3){15}}
\put(180,150){\line(-4,3){40}}
\multiput(140,180)(1.25,-3.75){20}{\circle*{1}}
\put(135,150){\circle*{3}}
\put(180,150){\circle*{3}}
\put(165,105){\circle*{3}}
\put(140,180){\circle*{3}}

\put(125,150){\makebox(0,0){\boldmath $\beta$}}
\put(190,150){\makebox(0,0){\boldmath $\alpha$}}
\put(170,95){\makebox(0,0){\boldmath $\gamma$}}
\put(135,190){\makebox(0,0){\boldmath $\delta$}}

\put(160,60){\makebox(0,0){\footnotesize{$s(\alpha,\beta,\gamma) = +1$}}}
\put(160,50){\makebox(0,0){\footnotesize{$s(\alpha,\beta,\delta) = -1$}}}

\put(240,150){\line(1,0){45}}
\put(240,150){\line(1,-6){5}}
\put(240,150){\line(2,3){30}}
\put(285,150){\line(-1,3){15}}
\put(285,150){\line(-4,-3){40}}
\multiput(245,120)(1.25,3.75){20}{\circle*{1}}
\put(240,150){\circle*{3}}
\put(285,150){\circle*{3}}
\put(270,195){\circle*{3}}
\put(245,120){\circle*{3}}

\put(230,150){\makebox(0,0){\boldmath $\beta$}}
\put(295,150){\makebox(0,0){\boldmath $\alpha$}}
\put(275,205){\makebox(0,0){\boldmath $\gamma$}}
\put(240,110){\makebox(0,0){\boldmath $\delta$}}

\put(265,60){\makebox(0,0){\footnotesize{$s(\alpha,\beta,\gamma) = -1$}}}
\put(265,50){\makebox(0,0){\footnotesize{$s(\alpha,\beta,\delta) = +1$}}}

\put(345,150){\line(1,0){45}}
\put(345,150){\line(1,6){5}}
\put(345,150){\line(2,3){30}}
\put(390,150){\line(-1,3){15}}
\put(390,150){\line(-4,3){40}}
\multiput(350,180)(2.5,1.5){10}{\circle*{1}}
\put(345,150){\circle*{3}}
\put(390,150){\circle*{3}}
\put(375,195){\circle*{3}}
\put(350,180){\circle*{3}}

\put(335,150){\makebox(0,0){\boldmath $\beta$}}
\put(400,150){\makebox(0,0){\boldmath $\alpha$}}
\put(380,205){\makebox(0,0){\boldmath $\gamma$}}
\put(340,190){\makebox(0,0){\boldmath $\delta$}}

\put(370,60){\makebox(0,0){\footnotesize{$s(\alpha,\beta,\gamma) = -1$}}}
\put(370,50){\makebox(0,0){\footnotesize{$s(\alpha,\beta,\delta) = -1$}}}

\put(55,30){\makebox(0,0){\footnotesize{(a)}}}
\put(160,30){\makebox(0,0){\footnotesize{(b)}}}
\put(265,30){\makebox(0,0){\footnotesize{(c)}}}
\put(370,30){\makebox(0,0){\footnotesize{(d)}}}

\put(0,10){\makebox(0,0)[l]{\footnotesize{
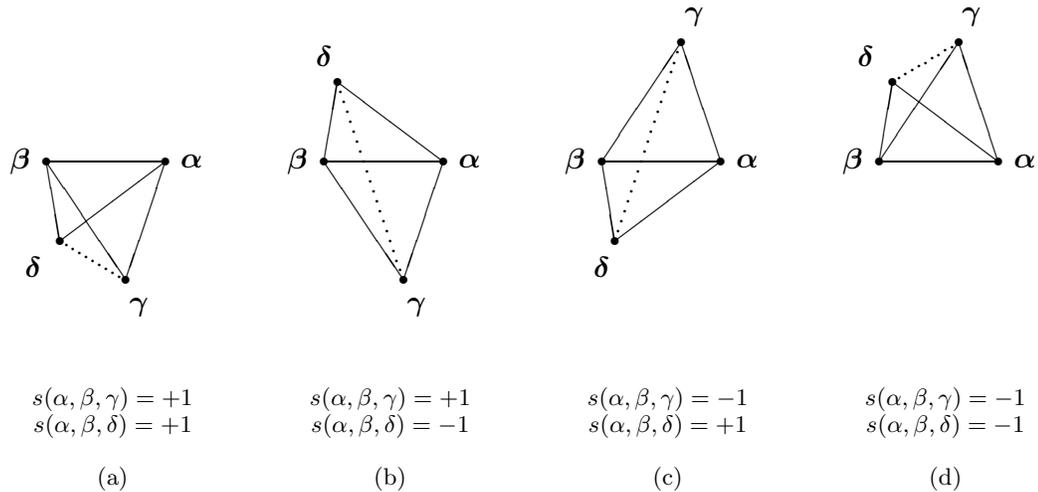
Fig. 3. The four possible orientations of $\gamma$ and $\delta$
relative to $\alpha$, $\beta$ represented schematically in ``plan view''
}}}
\put(0,0){\makebox(0,0)[l]{\footnotesize{
given the five angles
$\theta_{\alpha\beta}$,
$\theta_{\alpha\gamma}$,
$\theta_{\alpha\delta}$,
$\theta_{\beta\gamma}$,
$\theta_{\beta\delta}$,
represented by the unbroken lines. The angle
$\theta_{\gamma\delta}$ is represen-
}}}
\put(0,-10){\makebox(0,0)[l]{\footnotesize{
ted by the dotted line.
}}}

\end{picture}

Writing eqn. (\ref{2.16}) as a quadratic equation for
${\rm cos}\,\theta_{\gamma\delta}$
\be\label{2.17}
a\,{\rm cos}^{2}\theta_{\gamma\delta}
+ b\,{\rm cos}\,\theta_{\gamma\delta} + c
=
0
\ee
where the coefficients $a,b,c$ are functions of the other five angles,
the fact that ${\rm cos}\,\theta$ is monotonic for
$0 \leq \theta \leq \pi$ enables the solution to be written
\be\label{2.18}
{\rm cos}\,\theta_{\gamma\delta}
=
-\frac{b}{2a} +
s(\alpha,\beta,\gamma)\,s(\alpha,\beta,\delta)
\left|\frac{(b^{2} - 4ac)^{\frac{1}{2}}}{2a}\right|
\ee
i.e. the larger angle $\theta_{\gamma\delta}$ occurs when
\mbox{\boldmath $\gamma$} and \mbox{\boldmath $\delta$} are on
opposite sides of the plane formed by
\mbox{\boldmath $\alpha$} and \mbox{\boldmath $\beta$}, in which case
$s(\alpha,\beta,\gamma)\,s(\alpha,\beta,\delta) = -1$.

We now consider the number of independent continuous variables
$L(\alpha)$ and $L(\alpha,\beta)$ and discrete variables
$s(\alpha,\beta,\gamma)$. With $m$ matrices
$\alpha,\beta,\gamma\ldots$ there are $m$ associated vectors
\mbox{\boldmath $\alpha$}, \mbox{\boldmath $\beta$},
\mbox{\boldmath $\gamma$}$\ldots$.
The lengths of these vectors are given by the $m$ independent
variables $L(\alpha)$ via
$|\mbox{\boldmath $\alpha$}| = +(1 - L(\alpha)^{2})^{\frac{1}{2}}$.
Given the lengths of the vectors, the information given
by the variables $L(\alpha,\beta)$ is the set of angles
$\theta_{\alpha\beta}$. If some particular pair of vectors,
assumed non-parallel and of non-zero length, is
chosen to form a plane, say those labelled
{\bf 1} and {\bf 2}, then there are $m-2$
angles $\theta_{1\alpha}$ and $m-2$ angles $\theta_{2\alpha}$ required
to specify the orientation of the remaining $m-2$ vectors relative to
{\bf 1} and {\bf 2}, up to the reflectional
ambiguities. Including the angle $\theta_{12}$, this means that there
are $2m-3$ independent $\theta_{\alpha\beta}$, and hence,
given the $m$ $L(\alpha)$, $2m-3$ independent $L(\alpha,\beta)$.\footnote{
Alternatively, there are $\frac{1}{2}m(m-1)$ distinct $L(\alpha,\beta)$.
Setting $\alpha = 1$ and $\beta = 2$, there are $\frac{1}{2}(m-2)(m-3)$
independent equations of constraint (\ref{2.16}) for the sets of matrices
$\{1,2,\gamma,\delta\}$ (the left side of eqn. (\ref{2.16})
vanishes trivially if any two of the
$\alpha,\beta,\gamma,\delta$ are equal and is
invariant under any permutation of the $\alpha,\beta,\gamma,\delta$).
All other such eqns. (\ref{2.16}) not
involving the matrices labelled 1,2 are then dependent on these. Thus,
given the $m$ $L(\alpha)$,
there are $\frac{1}{2}m(m-1) - \frac{1}{2}(m-2)(m-3) = 2m-3$
independent $L(\alpha,\beta)$.} Thus
\be\label{2.19}
{\rm number\,\,of\,\,independent}\,\,L(\alpha),\,\,L(\alpha,\beta)
=
3m-3
\ee
Similarly, there are $m-2$ independent $s(1,2,\alpha)$ required to
specify the reflectional orientation of the $m-2$ vectors relative to
the plane formed by {\bf 1} and {\bf 2}. Thus
\be\label{2.20}
{\rm number\,\,of\,\,independent}\,\,s(\alpha,\beta,\gamma)
=
m-2
\ee

It should be noted that, while there are $3m$ continuous variables
associated with the $m$ $SU(2)$ matrices, there are only $3m-3$
independent continuous variables $L(\alpha),L(\alpha,\beta)$.
This is due to the fact that the
$L(\alpha)$ and $L(\alpha,\beta)$ are
invariant under any rotation of all $m$ vectors
\mbox{\boldmath $\alpha$}, \mbox{\boldmath $\beta$},
\mbox{\boldmath $\gamma$}$\ldots$; equivalently, the
${\rm Tr}(\alpha)$ and ${\rm Tr}(\alpha\beta)$
are invariant under a transformation
$\alpha \rightarrow g\alpha g^{-1}$, $g\,\in\,SU(2)$, of all $m$
matrices $\alpha,\beta,\gamma\ldots$. Thus the variables
$L(\alpha),L(\alpha,\beta)$ are unable to describe the three
``Euler angle'' variables
required to specify the overall orientation of the vectors
relative to some set of axes. However, together with the $m-2$
independent $s(\alpha,\beta,\gamma)$, they {\em are} fully sufficient
to describe any {\em trace} formed from the $m$ matrices, since such traces
are themselves invariant under such an overall transformation.

To summarize, using the formulae in this section, {\em any} trace
formed from the set of $m$ matrices $\alpha,\beta,\gamma\ldots$
can be fully expressed in
terms of some particular set of $3m-3$ independent continuous variables
$L(\alpha)$, $L(\alpha,\beta)$ (or alternatively
${\rm Tr}(\alpha)$, ${\rm Tr}(\alpha\beta)$) and $m-2$
independent discrete variables
$s(\alpha,\beta,\gamma)$. Thus the $SU(2)$ Mandelstam constaints may
be solved completely in terms of these variables.

{\bf 3. SU(2) lattice gauge theory}

\noindent
We now consider an $SU(2)$ pure gauge system on a $d$-dimensional
lattice with $n^{d}$ sites and with periodic boundary
conditions.\footnote{A set of arguments very similar to the following
can be given for the case of free boundary conditions.}
In the Lagrangian formulation, $d$ is the number of
space-time dimensions, while in the Hamiltonian formulation it is the
number of space dimensions, time remaining continuous.
The number $3\cal{N}$ of independent degrees of freedom
of either lattice system
is given by the dimension of the quotient space
$\otimes_{{\rm links}}SU(2)/\otimes_{{\rm sites}}SU(2)$ i.e.
\bea\label{3.1}
3\cal{N}
&=&
{\rm dim}\,\,SU(2)\times ( {\rm number\,\,of\,\,links}
- {\rm number\,\,of\,\,sites} ) \\
\label{3.2}
&=&
3(d-1)n^{d}
\eea

A gauge in which the gauge
redundancy is reduced to the greatest possible extent is given by
making local gauge transformations such that as many links as
possible are fixed to the unit matrix $I$. Such a
configuration of fixed links defines a ``maximal tree'', characterized
by the fact that, while the maximum number of links have been fixed,
it is impossible to have any of them form a loop. An
example of a maximal tree for $d=2$, $n=4$ is shown in fig. 4. The
number of unfixed links on a maximal tree for any $d$ is ${\cal N}+1$.
The ``extra'' three degrees of freedom coming from the unfixed
``${\cal N}+1$'th'' link are due to the fact that it is still possible to
make a global gauge transformation, rotating the overall system
of $SU(2)$ matrices' associated vectors relative to
some frame of reference. Specifying this relative orientation requires
three Euler angles, so that if done this would account for these three
degrees of freedom.

\noindent
\begin{picture}(425,180)(0,-10)

\put(175,50){\line(0,1){80}}
\put(195,50){\line(0,1){80}}
\put(215,50){\line(0,1){80}}
\put(235,50){\line(0,1){80}}
\put(175,50){\line(1,0){80}}
\put(175,70){\line(1,0){80}}
\put(175,90){\line(1,0){80}}
\put(175,110){\line(1,0){80}}

\multiput(175,50)(0,0.5){121}{\circle*{2.0}}
\multiput(195,50)(0,0.5){121}{\circle*{2.0}}
\multiput(215,50)(0,0.5){121}{\circle*{2.0}}
\multiput(235,50)(0,0.5){121}{\circle*{2.0}}
\multiput(175,50)(0.5,0){121}{\circle*{2.0}}

\put(0,10){\makebox(0,0)[l]{\footnotesize{
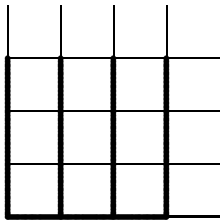
Fig. 4. A possible maximal gauge fixing tree for a
$4\!\times\! 4$ lattice with periodic boundary conditions. Fixed
}}}
\put(0,0){\makebox(0,0)[l]{\footnotesize{
links are indicated by the heavy lines.
}}}

\end{picture}

The gauge invariant and hence ``physical''
quantities are the traces of loops of link matrices.
If (and it is emphasized that this is {\em not} what will in fact be
done) the ${\cal N}+1$ unfixed links on the maximally gauge fixed lattice
were represented by the $m$ matrices
$\alpha,\beta,\gamma\ldots$, so that $m = {\cal N}+1$, then
from eqns. (\ref{2.19}) and (\ref{2.20})
\bea\label{3.3}
{\rm number\,\,of\,\,independent}
\,\,{\rm Tr}(\alpha),\,\,{\rm Tr}(\alpha\beta)
&=&
3{\cal N} \\
{\rm number\,\,of\,\,independent}
\,\,s(\alpha,\beta,\gamma)
&=&
{\cal N}-1
\eea
Thus it is seen that when dealing only with traces
and expressing them in terms of independent
${\rm Tr}(\alpha),{\rm Tr}(\alpha\beta)$ and
$s(\alpha,\beta,\gamma)$ as described in section 2, the number
of continuous degrees of freedom is exactly correct.

This demonstration that for physical quantities the number of
independent ${\rm Tr}(\alpha)$ and ${\rm Tr}(\alpha\beta)$
is correct involves some particular maximal gauge fixing -- the
${\rm Tr}(\alpha),{\rm Tr}(\alpha\beta)$ and
$s(\alpha,\beta,\gamma)$ used above
are certainly not gauge invariant quantities.
In order to arrive at a fully
gauge invariant formulation of the $SU(2)$ theory, rather than taking
the $m = {\cal N}+1$ matrices $\alpha,\beta,\gamma\ldots$
to be the remaining links after
some maximal gauge fixing, it is necessary to construct from the
lattice links a different set of ${\cal N}+1$ independent
matrices without reference to any gauge fixing.
This set must satisfy two
criteria: firstly, the trace of each of the matrices must be gauge
invariant i.e. the matrices must
each represent a loop; secondly, it must be possible to
construct from these matrices
{\em any} other loop matrix by taking some appropriate
product of matrices and their inverses from the set. Together, these
two criteria imply that the matrices must each consist of products of
links which start and end at the same (arbitrary) point on the
lattice. Furthermore, the second criterion implies that, in $d$
dimensions, $d$ of these matrices must consist of products of links
which ``wrap around'' the periodic lattice, one in each of the
directions. When traced, these particular matrices give Polyakov loops.
An example of a set of such matrices
for $d=2$, $n=3$ (hence ${\cal N}+1 = 10$) is shown in fig. 5.

\noindent
\begin{picture}(425,300)

\put(5,200){\line(0,1){60}}
\put(25,200){\line(0,1){60}}
\put(45,200){\line(0,1){60}}
\put(5,200){\line(1,0){60}}
\put(5,220){\line(1,0){60}}
\put(5,240){\line(1,0){60}}
\multiput(5,200)(0.5,0){121}{\circle*{1.8}}
\multiput(40,200)(-0.5,0.5){7}{\circle*{1}}
\multiput(40,200)(-0.5,-0.5){7}{\circle*{1}}

\put(30,180){\makebox(0,0)[l]{$U_{1}$}}

\put(95,200){\line(0,1){60}}
\put(115,200){\line(0,1){60}}
\put(135,200){\line(0,1){60}}
\put(95,200){\line(1,0){60}}
\put(95,220){\line(1,0){60}}
\put(95,240){\line(1,0){60}}
\multiput(95,200)(0,0.5){121}{\circle*{1.8}}
\multiput(95,235)(-0.5,-0.5){7}{\circle*{1}}
\multiput(95,235)(0.5,-0.5){7}{\circle*{1}}

\put(120,180){\makebox(0,0)[l]{$U_{2}$}}

\put(185,200){\line(0,1){60}}
\put(205,200){\line(0,1){60}}
\put(225,200){\line(0,1){60}}
\put(185,200){\line(1,0){60}}
\put(185,220){\line(1,0){60}}
\put(185,240){\line(1,0){60}}
\multiput(185,200)(0.5,0){41}{\circle*{1.8}}
\multiput(205,200)(0,0.5){41}{\circle*{1.8}}
\multiput(185,220)(0.5,0){41}{\circle*{1.8}}
\multiput(185,203)(0,0.5){35}{\circle*{1.8}}
\multiput(200,200)(-0.5,0.5){7}{\circle*{1}}
\multiput(200,200)(-0.5,-0.5){7}{\circle*{1}}

\put(210,180){\makebox(0,0)[l]{$U_{3}$}}

\put(275,200){\line(0,1){60}}
\put(295,200){\line(0,1){60}}
\put(315,200){\line(0,1){60}}
\put(275,200){\line(1,0){60}}
\put(275,220){\line(1,0){60}}
\put(275,240){\line(1,0){60}}
\multiput(275,200)(0.5,0){81}{\circle*{1.8}}
\multiput(315,200)(0,0.5){41}{\circle*{1.8}}
\multiput(295,220)(0.5,0){41}{\circle*{1.8}}
\multiput(295,203)(0,0.5){35}{\circle*{1.8}}
\multiput(275,203)(0.5,0){41}{\circle*{1.8}}
\multiput(310,200)(-0.5,0.5){7}{\circle*{1}}
\multiput(310,200)(-0.5,-0.5){7}{\circle*{1}}

\put(300,180){\makebox(0,0)[l]{$U_{4}$}}

\put(365,200){\line(0,1){60}}
\put(385,200){\line(0,1){60}}
\put(405,200){\line(0,1){60}}
\put(365,200){\line(1,0){60}}
\put(365,220){\line(1,0){60}}
\put(365,240){\line(1,0){60}}
\multiput(365,200)(0.5,0){121}{\circle*{1.8}}
\multiput(425,200)(0,0.5){41}{\circle*{1.8}}
\multiput(405,220)(0.5,0){41}{\circle*{1.8}}
\multiput(405,203)(0,0.5){35}{\circle*{1.8}}
\multiput(365,203)(0.5,0){81}{\circle*{1.8}}
\multiput(420,200)(-0.5,0.5){7}{\circle*{1}}
\multiput(420,200)(-0.5,-0.5){7}{\circle*{1}}

\put(390,180){\makebox(0,0)[l]{$U_{5}$}}

\put(5,80){\line(0,1){60}}
\put(25,80){\line(0,1){60}}
\put(45,80){\line(0,1){60}}
\put(5,80){\line(1,0){60}}
\put(5,100){\line(1,0){60}}
\put(5,120){\line(1,0){60}}
\multiput(5,80)(0,0.5){81}{\circle*{1.8}}
\multiput(5,120)(0.5,0){41}{\circle*{1.8}}
\multiput(25,100)(0,0.5){41}{\circle*{1.8}}
\multiput(8,100)(0.5,0){35}{\circle*{1.8}}
\multiput(8,80)(0,0.5){41}{\circle*{1.8}}
\multiput(20,100)(-0.5,0.5){7}{\circle*{1}}
\multiput(20,100)(-0.5,-0.5){7}{\circle*{1}}

\put(30,60){\makebox(0,0)[l]{$U_{6}$}}

\put(95,80){\line(0,1){60}}
\put(115,80){\line(0,1){60}}
\put(135,80){\line(0,1){60}}
\put(95,80){\line(1,0){60}}
\put(95,100){\line(1,0){60}}
\put(95,120){\line(1,0){60}}
\multiput(95,83)(0.5,0){41}{\circle*{1.8}}
\multiput(115,83)(0,0.5){75}{\circle*{1.8}}
\multiput(115,120)(0.5,0){41}{\circle*{1.8}}
\multiput(135,100)(0,0.5){41}{\circle*{1.8}}
\multiput(118,100)(0.5,0){35}{\circle*{1.8}}
\multiput(118,80)(0,0.5){41}{\circle*{1.8}}
\multiput(95,80)(0.5,0){47}{\circle*{1.8}}
\multiput(130,100)(-0.5,0.5){7}{\circle*{1}}
\multiput(130,100)(-0.5,-0.5){7}{\circle*{1}}

\put(120,60){\makebox(0,0)[l]{$U_{7}$}}

\put(185,80){\line(0,1){60}}
\put(205,80){\line(0,1){60}}
\put(225,80){\line(0,1){60}}
\put(185,80){\line(1,0){60}}
\put(185,100){\line(1,0){60}}
\put(185,120){\line(1,0){60}}
\multiput(185,83)(0.5,0){81}{\circle*{1.8}}
\multiput(225,83)(0,0.5){75}{\circle*{1.8}}
\multiput(225,120)(0.5,0){41}{\circle*{1.8}}
\multiput(245,100)(0,0.5){41}{\circle*{1.8}}
\multiput(228,100)(0.5,0){35}{\circle*{1.8}}
\multiput(228,80)(0,0.5){41}{\circle*{1.8}}
\multiput(185,80)(0.5,0){87}{\circle*{1.8}}
\multiput(240,100)(-0.5,0.5){7}{\circle*{1}}
\multiput(240,100)(-0.5,-0.5){7}{\circle*{1}}

\put(210,60){\makebox(0,0)[l]{$U_{8}$}}

\put(275,80){\line(0,1){60}}
\put(295,80){\line(0,1){60}}
\put(315,80){\line(0,1){60}}
\put(275,80){\line(1,0){60}}
\put(275,100){\line(1,0){60}}
\put(275,120){\line(1,0){60}}
\multiput(275,80)(0,0.5){121}{\circle*{1.8}}
\multiput(275,140)(0.5,0){41}{\circle*{1.8}}
\multiput(295,120)(0,0.5){41}{\circle*{1.8}}
\multiput(278,120)(0.5,0){35}{\circle*{1.8}}
\multiput(278,80)(0,0.5){81}{\circle*{1.8}}
\multiput(290,120)(-0.5,0.5){7}{\circle*{1}}
\multiput(290,120)(-0.5,-0.5){7}{\circle*{1}}

\put(300,60){\makebox(0,0)[l]{$U_{9}$}}

\put(365,80){\line(0,1){60}}
\put(385,80){\line(0,1){60}}
\put(405,80){\line(0,1){60}}
\put(365,80){\line(1,0){60}}
\put(365,100){\line(1,0){60}}
\put(365,120){\line(1,0){60}}
\multiput(365,83)(0.5,0){41}{\circle*{1.8}}
\multiput(385,83)(0,0.5){115}{\circle*{1.8}}
\multiput(385,140)(0.5,0){41}{\circle*{1.8}}
\multiput(405,120)(0,0.5){41}{\circle*{1.8}}
\multiput(388,120)(0.5,0){35}{\circle*{1.8}}
\multiput(388,80)(0,0.5){81}{\circle*{1.8}}
\multiput(365,80)(0.5,0){47}{\circle*{1.8}}
\multiput(400,120)(-0.5,0.5){7}{\circle*{1}}
\multiput(400,120)(-0.5,-0.5){7}{\circle*{1}}

\put(390,60){\makebox(0,0)[l]{$U_{10}$}}

\put(0,20){\makebox(0,0)[l]{\footnotesize{
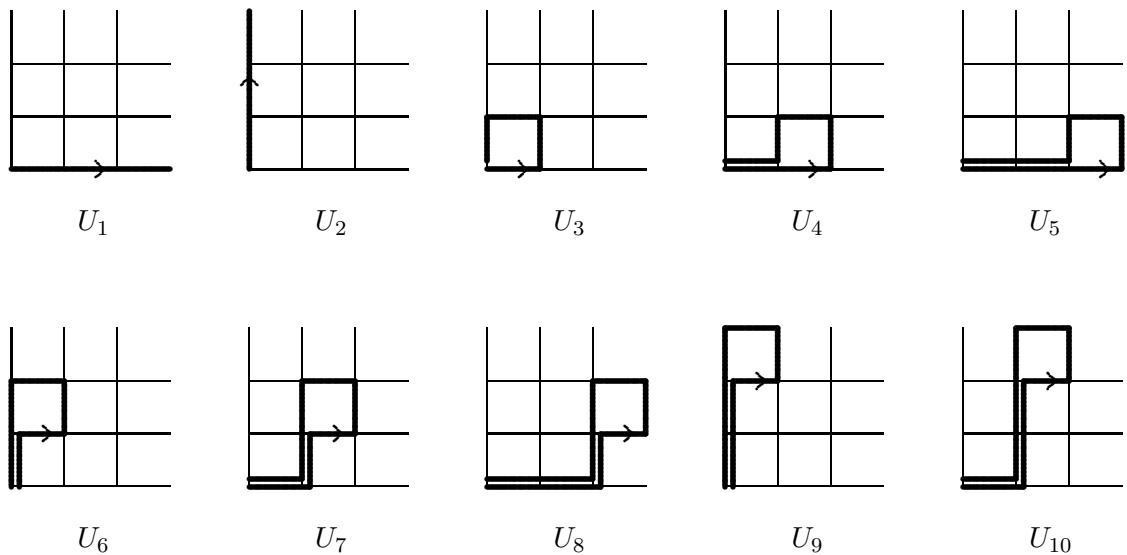
Fig. 5. A possible set of 10 basis matrices, indicated
by the heavy lines, for a $3\!\times\! 3$ lattice with periodic
}}}
\put(0,10){\makebox(0,0)[l]{\footnotesize{
boundary conditions.
}}}

\end{picture}

It can be seen in this example that, indeed, any other loop matrix can
be constructed from the matrices shown.
For $d = 2$ it is easy to see how this particular choice of matrices
(which is only one among an infinite number) can be
extended for lattices of other sizes. For $d = 3,4$, analogous sets can
be constructed, though not so easily.

As the matrices all start and end at the same point on the periodic
lattice, it is clear that the trace of any product of these matrices
and their inverses
is gauge invariant. In particular, this obviously includes the
${\rm Tr}(\alpha)$ and ${\rm Tr}(\alpha\beta)$. Crucially, this also means
that the discrete variables $s(\alpha,\beta,\gamma)$ too are
manifestly gauge invariant. For example, for the choice of
matrices shown in fig. 5, the variable $s(4,7,10)$ is given from
eqn. (\ref{2.15}) by
\be\label{3.4}
s(4,7,10)
=
\frac{L(4,7,10)}{|L(4,7,10)|}
\ee
with, using the original definition eqn. (\ref{2.6}),

\noindent
\addtocounter{equation}{1}
\begin{picture}(424,95)(0,5)

\put(74,50){\makebox(0,0){${\displaystyle L(4,7,10) = -\frac{1}{2}}$}}
\put(120,20){\framebox(20,60){}}
\put(165,50){\makebox(0,0){${\displaystyle +\frac{1}{4}}$}}
\put(180,20){\framebox(20,19){}}
\put(180,41){\framebox(20,39){}}
\put(225,50){\makebox(0,0){${\displaystyle +\frac{1}{4}}$}}
\put(240,20){\line(1,0){20}}
\put(240,20){\line(0,1){60}}
\put(260,20){\line(0,1){19}}
\put(242,39){\line(1,0){18}}
\put(242,39){\line(0,1){22}}
\put(242,61){\line(1,0){18}}
\put(260,61){\line(0,1){19}}
\put(240,80){\line(1,0){20}}
\put(244,41){\framebox(16,18){}}
\put(285,50){\makebox(0,0){${\displaystyle +\frac{1}{4}}$}}
\put(300,20){\framebox(20,39){}}
\put(300,61){\framebox(20,19){}}
\put(345,50){\makebox(0,0){${\displaystyle -\frac{1}{4}}$}}
\put(360,20){\framebox(20,19){}}
\put(360,41){\framebox(20,18){}}
\put(360,61){\framebox(20,19){}}
\put(424,50){\makebox(0,0)[r]{$(29)$}}

\end{picture}

\noindent
(the position of the above loops on the lattice of fig. 5 is not
indicated but should be clear).

Given some $SU(2)$ lattice gauge system, it is therefore necessary
to choose some basis set of ${\cal N}+1$ independent matrices
$\alpha,\beta,\gamma\ldots$ satisfying the above two criteria
and then from these to choose some
set of $2{\cal N}-1$ independent ${\rm Tr}(\alpha\beta)$ and
${\cal N}-1$ independent $s(\alpha,\beta,\gamma)$, the ${\cal N}+1$
${\rm Tr}(\alpha)$ being immediately settled. In the example of fig. 5,
two of the ${\rm Tr}(\alpha)$ are the Polyakov loops while the
remaining eight are single plaquette Wilson loops located at eight of
the nine elementary plaquettes of the lattice. It is immediately clear
that this choice of basis matrices does not give a set of variables
which is invariant under translation, $90^{\rm o}$ rotation and
reflection.
Because, for the discrete variables, ${\cal N}-1 = (d-1)n^{d} - 1$
is not proportional to the total number $n^{d}$ of lattice sites, it
is in fact impossible to construct a set of variables
which has the same spatial symmetries as the lattice.

{\bf 4. Conclusions}

\noindent
It has been shown here how an $SU(2)$ lattice gauge theory may be
fully described in terms of the correct number $3{\cal N}$ of gauge
invariant continuous loop variables together with ${\cal N}-1$
discrete $\pm 1$ variables.
Apart from the question of quantization,
two practical possibilities immediately arise.
The first concerns Monte Carlo simulations in the Lagrangian
formulation. Instead of writing the action in terms of the individual
link variables, it could be written in terms of a set of gauge
invariant variables. The algebraic expression for $S$ would be
extremely complicated, but the fact that it would (naturally)
involve discrete
$\pm 1$ variables might enable efficient updating, similar to the
extremely efficient Wolff embedding algorithms used for
non-linear $O(N)$ $\sigma$-models \cite{wolff1}\cite{wolff2}\cite{wolff3}.
The second possibility concerns (semi-)analytic approaches in the
Hamiltonian formulation. In the Kogut-Susskind Hamiltonian,
in Schr\"odinger representation the
electric field operators are differential operators, so that with the
wavefunction $\psi$ a function of a set of gauge invariant variables,
the Schr\"odinger equation $H\psi = E\psi$ is a partial differential
eigenvalue equation in these variables. On a finite lattice, such an
equation might be amenable to numerical solution; alternatively
$\psi$ could be expanded as a Taylor series in these variables. This
latter, but on an effectively infinite lattice,
is very similar to the approach advocated in ref. \cite{chlsnjw}.

{\bf Acknowledgements}

\noindent
I wish to thank Chris Korthals-Altes, Chris Llewellyn Smith and Stam
Nicolis for many useful discussions. The financial support of a
SERC/NATO Postdoctoral Fellowship is gratefully acknowledged.


\end{document}